\documentclass{iau} 
\usepackage{graphicx}

\usepackage{amsmath}
\usepackage{amssymb,latexsym,mathrsfs, bm}
\usepackage{color}
\usepackage{float}

\title[Neutral hydrogen in the post-reionization universe ] 
{Neutral hydrogen in the post-reionization universe}

\author[Hamsa Padmanabhan]   
{Hamsa Padmanabhan
 }

\affiliation{ETH Zurich, \\ Wolfgang-Pauli-Strasse 27,
CH 8093 Zurich, Switzerland \\ email: {\tt hamsa.padmanabhan@phys.ethz.ch} \\}

\pubyear{2018}
\volume{333}  
\jname{Peering towards Cosmic Dawn}
\editors{Vibor Jeli\'c \& Thijs van der Hulst, eds.}
\begin{document}

\maketitle

\begin{abstract}
The evolution of neutral hydrogen (HI) across redshifts is a powerful
probe of cosmology, large scale structure in the universe and the
intergalactic medium. Using a data-driven halo model to describe the
distribution of HI in the post-reionization universe ($z \sim $ 5 to 0),  we
obtain the best-fitting parameters from a rich sample of observational
data: low redshift 21-cm emission line studies, intermediate redshift
intensity mapping experiments, and higher redshift Damped Lyman Alpha
(DLA) observations.  Our model describes the abundance and clustering of
neutral hydrogen across redshifts 0 - 5, and is useful for investigating
different aspects of galaxy evolution and for comparison with
hydrodynamical simulations. The framework can be applied for forecasting future
observations with neutral hydrogen, and extended to the case of intensity mapping with molecular and other line transitions  at intermediate redshifts.

\end{abstract}

\firstsection 

\section{Introduction}
Hydrogen is the most abundant element in the universe, hence mapping the evolution of neutral hydrogen across cosmic time promises valuable constraints on galaxy evolution, theories of gravity and fundamental physics.
There are many ongoing and planned measurements which will give us clues to the distribution and evolution of neutral hydrogen (hereafter HI) over the last 12 billion years (redshifts 0 to 5). In the post-reionization epoch ($z \lesssim 6$), the 21-cm line emission of HI is expected to act as a good tracer of the underlying dark matter distribution, due to the absence of the complex reionization astrophysics. This is not necessarily the case prior to and during reionization, because the neutral hydrogen exists both within and outside galaxies as the ionized regions form and overlap.  Hence, while the epochs prior to redshift 6 ($z \ge 6$) are interesting for reionization studies, it is useful to use the observations of post-reionization HI to investigate large-scale structure.  Neutral hydrogen at these times exists in the form of dense clumps in galaxies, Lyman-limit systems and Damped Lyman-Alpha systems (DLAs). At the lower end of the redshift range ($z \sim 0-1$), 21-cm emission line surveys of galaxies have been used as probes of the neutral hydrogen distribution (e.g., Zwaan et al. 2005, Martin et al. 2012). The limits of current radio facilities, however, hamper detection of 21-cm in emission above redshifts of $z \sim 0.1$ from normal galaxies. At higher redshifts ($z > 2$), the HI distribution has been  studied via the identification of Damped Lyman-Alpha systems (DLAs) in absorption line surveys, e.g. Noterdaeme et al. (2012), Crighton et al. (2015) --- which are believed to be the primary reservoirs of neutral hydrogen at these epochs.  

A relatively new technique used to study HI evolution in the post-reionization universe is  \textit{intensity mapping}. In this  technique, the large-scale distribution of a tracer (like HI) is mapped without resolving the individual galaxies which host the tracer.  Being faster and less expensive than galaxy surveys, intensity mapping had already been shown to provide constraints on the  HI  abundances and clustering at moderate redshifts, competitive with those from next generation experiments. Early work investigating these aspects included, e.g.  Bharadwaj et al. (2009), Wyithe \& Loeb (2008).
The first HI intensity mapping detection used cross-correlation measurements with the DEEP2 optical galaxy redshift survey  (Chang et al. 2010), which has been followed up since then resulting in tighter constraints on the abundance and clustering of HI. Several efforts with future and current facilities are underway to map the distribution of HI, using the CHIME, BINGO, TianLai, HIRAX, FAST, MeerKAT, ASKAP and SKA experiments, among others.

\section{A halo model for cosmological neutral hydrogen}

On the theoretical front, a number of efforts have been made towards modelling the 21 cm signal to be observed with current and future experiments -- both at high and low redshifts. Semi-analytical models and hydrodynamical simulations (e.g., Dave et al. 2013) have focused on reproducing the HI content of galaxies and the HI-stellar scaling relations at low redshifts. The simulations also involve modelling the observed DLA properties at high redshifts (e.g., Bird et al. 2014; Rahmati et al. 2013; Rahmati \& Schaye 2014) and in the presence of strong stellar feedback, can reproduce the observed abundances and clustering [e.g., Font-Ribera et al. (2012)\footnote{This measurement was recently updated in P{\'e}rez-R{\`a}fols et al.(2018).}] of DLAs, but may lead to an excess of HI abundances at low redshifts.

Analytical techniques offer complementary insights into the processes governing the HI content of galaxies. The 21 cm intensity mapping observables, particularly the HI bias and power spectrum (e.g., Marin  et al. 2010) as well as DLA properties (e.g., Barnes \& Haehnelt 2014) are modelled using prescriptions for assigning HI mass to dark matter halos. In Padmanabhan et al. (2016), the 21-cm  and DLA-based analytical approaches are combined towards building a consistent model of HI evolution across redshifts. It is found that a model that reproduces both the abundances of low redshift systems, and the DLA clustering at high redshifts required a fairly rapid transition of HI from low- to high-mass haloes from $z \sim 0-3$. This is also borne out by a complementary, abundance matching approach (Padmanabhan \& Kulkarni 2017),  which has the advantage of being completely empirical and consistent with the stellar-gas scaling relations found in simulations and observations. This study also provides a measure of the level of uncertainty in the data so far, and the resultant scatter in the high redshift HI - halo mass relation.

In the light of the presently available constraints, it is important to formulate a framework for combining the data into an evolutionary sequence which smoothly translates between the high- and low-redshift observables. As has been shown both for the case of dark matter (e.g., Sheth \& Tormen 2002) and for the case of stellar evolution (e.g., Moster et al. 2013), the \textit{halo model framework} is very well suited to capture the abundance and clustering of systems especially on non-linear scales. It is also a useful tool for investigating the relations between stellar and gas evolution in galaxies, and for comparison to hydrodynamical simulations. It is possible to develop a halo model framework for cosmological neutral hydrogen observables (e.g., Padmanabhan \& Refregier 2017, Padmanabhan, Refregier \& Amara 2017). To do this, the datasets from the 21 cm emission, intensity mapping and DLA observations are combined using a Markov Chain Monte Carlo (MCMC) analysis. Two main ingredients are found to describe the HI observations: (i) the HI - halo mass relation and (ii) the HI profile.

\subsection{The HI - halo mass relation}
The HI-halo mass relation (HIHM) denoted by $M_{\rm HI} (M,z)$, describes the average HI content of a halo of mass $M$ at redshift $z$.
Several approaches in the literature have focused on modelling this function, e.g.,  Villaescusa-Navarro et al. (2014), Barnes \& Haehnelt (2014), Bagla et al. (2010). 
We find it useful to describe it by the form:
\begin{eqnarray}
M_{\rm HI} (M) &=& \alpha f_{H,c} M \left(\frac{M}{10^{11} h^{-1} M_{\odot}}\right)^{\beta} \exp\left[-\left(\frac{v_{c0}}{v_c(M)}\right)^3\right] \nonumber \\
\end{eqnarray}
The relation involves three free parameters, $\alpha$, $\beta$ and $v_{c,0}$:
(i) $\alpha$ is an overall normalization which represents the fraction of HI, relative to cosmic ($f_{\rm H,c}$) associated with a dark matter halo of mass $M$. (ii) $\beta$ is the logarithmic slope of the  relation, and was set to unity in some of the previous analyses. However, the best-fitting value of $\beta$ was found to be less than unity so as to fit the observations of the HI mass function at $z \sim 0$ (Padmanabhan \& Refregier 2017).  
(iii) The parameter $v_{c,0}$ represents the minimum virial velocity of a halo able to host HI. In previous studies, this value was set to 30 km/s at low redshifts and increased to $\sim 50$ km/s to fit the DLA data at higher redshifts. 

\subsection{The HI profile}

\begin{figure*}
\begin{center}
\includegraphics[scale=0.5]{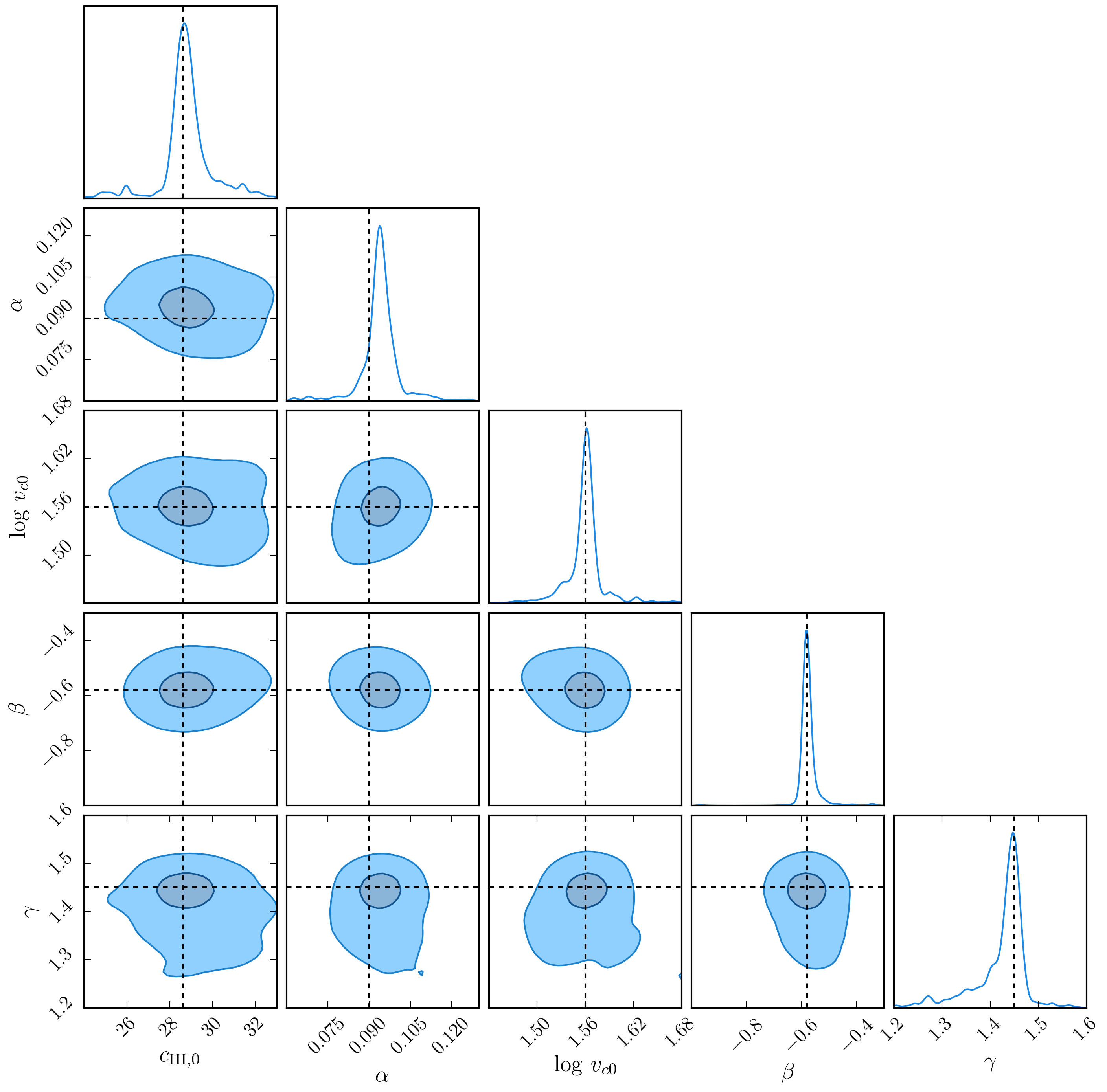}
\end{center}
\caption{Parameter space showing the constraints from the MCMC analysis for the HI halo model. The contours indicate 1- and 2$\sigma$ confidence levels. Crosshairs indicate the maximum likelihood estimate of the parameters (the best-fitting value). Marginal distributions of each parameter are shown in the diagonal panels. [Figure from Padmanabhan, Refregier \& Amara (2017).] }
\label{fig:cornerplot}
\end{figure*}

This describes the distribution of HI in a dark matter halo of mass $M$, as a function of radial distance $r$ from the centre of the halo. The form of the profile suggested by low redshift observations (e.g., Wang et al. 2014, Bigiel \& Blitz 2012) is exponential, which can be expressed as:

\begin{equation}
\rho(r,M) = \rho_0 \exp(-r/r_s)
\label{rhodefexp}
\end{equation}
where $r_s$ is the scale radius of the host halo, which is expressible in terms of the concentration parameter and the virial radius, as:
\begin{equation}
 c_{\rm HI}(M,z) =  c_{\rm HI, 0} \left(\frac{M}{10^{11} M_{\odot}} \right)^{-0.109} \frac{4}{(1+z)^{\gamma}}.
\end{equation} 
This introduces the two parameters: the overall normalization $c_{\rm HI, 0}$, and the slope $\gamma$ which describes the redshift evolution. We thus see that the HI-related quantities can be described by a total of five free parameters.\footnote{An altered NFW profile was found to be favored in some studies of high redshift DLA systems [e.g., Barnes \& Haehnelt  (2014)], and is also explored in Padmanabhan, Refregier \& Amara (2017).} 
An MCMC analysis fitted to all the data and using the exponential profile, leads to the results  in Fig. \ref{fig:cornerplot}.

\subsection{Insights and consistency with other observations}
The halo model framework  leads to the following best fitting values for the free parameters: $c_{\rm HI, 0} = 28.65 \pm 1.76, \alpha = 0.09 \pm 0.01, \log v_{\rm c,0} = 1.56 \pm 0.04, \beta = -0.58 \pm 0.06, \gamma = 1.45 \pm 0.04$. The mean values provide realistic estimates (in the light of all the current data) of the evolution of the HI-related quantities, and the error bars chiefly encompass the statistical uncertainties on the parameters. Apart from being consistent with the present set of observables, the framework also provides physical insights into the population of HI systems, and their evolution with cosmic time:

\begin{itemize}
\item The non-unity slope of the HI-halo mass relation is found to lead to a better match to the HI mass function (especially for large halo masses, at low redshifts) -- hence,  the low redshift HI mass function data chiefly drives the value of $\beta$.

\item The exponential profile is found to be a good fit to the low redshift observations -- which typically have the least statistical uncertainties, and is also  a good fit to the high-redshift DLA data. If only the current high-redshift data is considered, both the exponential and altered NFW profiles lead to equally good fits. 
\end{itemize}

\begin{figure}
  \begin{center}
    \includegraphics[scale=0.37]{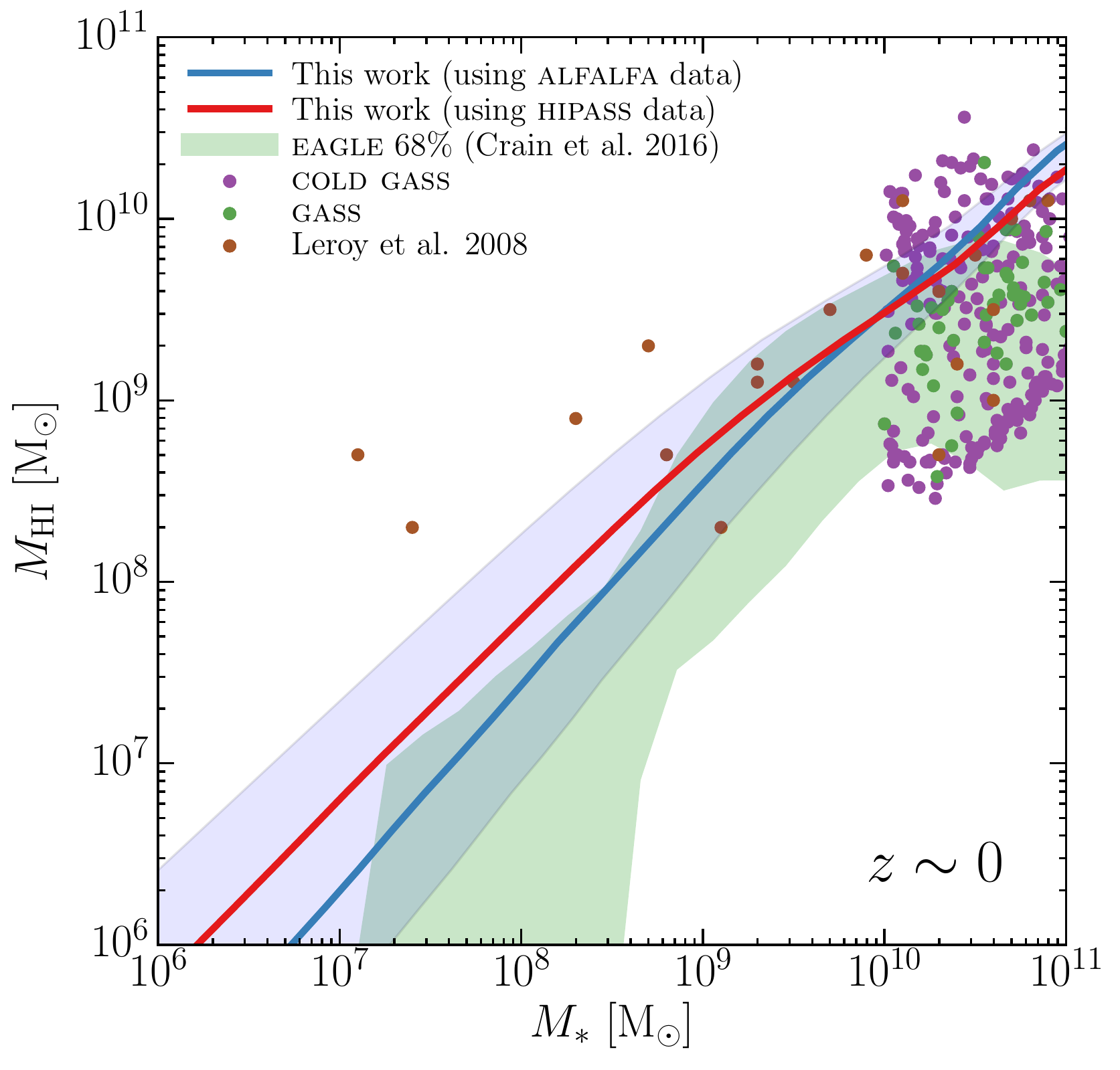} \includegraphics[scale=0.34]{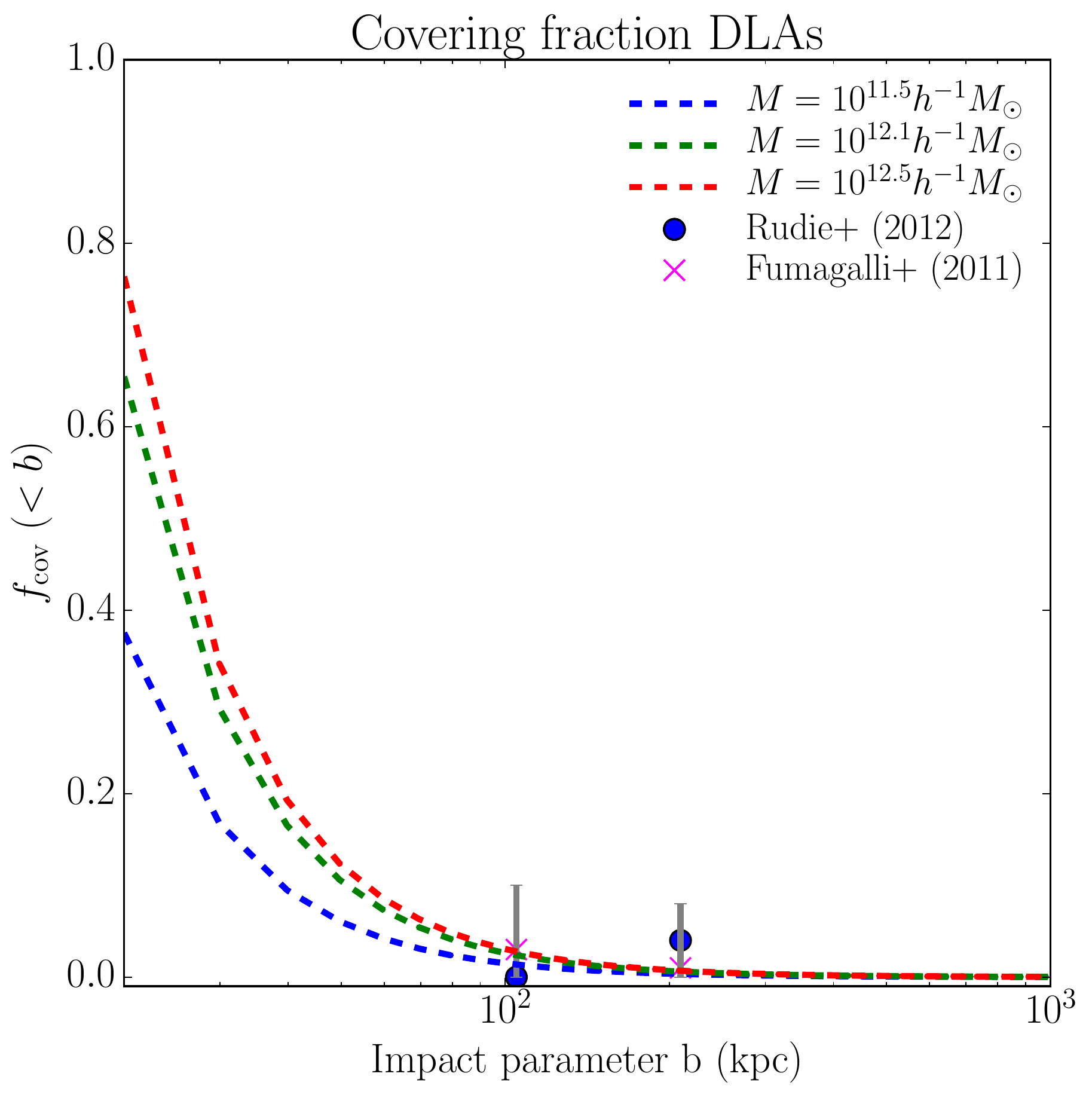} 
  \end{center}
  \caption{\textit{Left panel:} Solid curves show the HI-mass stellar-mass relation at $z \sim 0$ from HI abundance
    matching (blue band shows 68\% scatter) combined with the stellar-halo mass (SHM) relation from
   Moster et al. (2013).  The data from the EAGLE simulations (green band; Crain et al. 2017) and individual objects
    in various surveys (Leroy et al. 2008, Catinella et al. 2013, Saintonge et al. 2011) are also shown for comparison. \textit{Right panel:} The DLA covering fractions at $z \sim 2.5$ as a function of impact parameter from observations and simulations (Rudie et al. 2012, Fumagalli et al. 2011), along  with the model predictions for a range of host halo masses.  [Figures from Padmanabhan \& Kulkarni (2017) and Padmanabhan, Refregier \& Amara (2017).]}
  \label{fig:mhimstar}
\end{figure}

These findings can be connected to  other observations of low redshift stellar-gas scaling relations and DLA covering fraction observations. Examples are shown in Fig. \ref{fig:mhimstar}, where the left panel indicates the HI mass - stellar mass relation obtained by abundance matching, and the right panel indicates the DLA covering fraction observations, and the model predictions for a representative range of host halo masses. It is found that both these sets of observations are consistent with the model (even though the model was calibrated against a different set of data). Further, this framework leads to surface density profiles at low redshifts which are fully consistent with observations, e.g., in Bigiel \& Blitz (2012), which serves as an additional, independent  validation of the profile parameters.

\section{Summary and outlook}

We have seen that the various observations of neutral hydrogen in the post-reionization universe can be combined into a halo model framework, and the best-fitting parameters -- both for the HI-halo mass relation and the HI profile -- can be obtained by a Bayesian MCMC analysis. This framework is also consistent with  a number of complementary high- and low-redshift observations of neutral hydrogen from galaxies and DLAs. It has a wide range of  potential applications:

\begin{itemize}
\item A preliminary analysis of current HI data showed (Padmanabhan, Choudhury \& Refregier 2014) that the present astrophysical constraints lead to about 60\% - 100\% overall uncertainty in the estimation of the HI power spectrum. This `astrophysical systematic' can be more realistically quantified with the halo model framework (Padmanabhan, Refregier and Amara, in preparation). With cross-correlation data already becoming available (e.g. Switzer et al. 2013, Anderson et al. 2017), there is scope for introducing further parameters into the model and exploring non-linear scales in more detail. 

\item A straightforward extension of these methods to other probes [e.g., the case for the carbon monoxide (CO) molecule 1-0 transition is explored in Padmanabhan (2017)] is useful for exploiting the potential of cross-correlating intensity maps in the post-reionization universe -- to better understand the baryon cycle and stellar-cold gas relations. Further, comparison of the results to galaxy formation simulations will shed light on the physical processes governing feedback and outflows.

\item 
 It is also of interest to explore the possibility of halo model techniques (possibly with additional parameters) to achieve the `astrophysical separation' which is required for studies of the Epoch of Reionization. An ambitious future goal would be to use the isolation of astrophysical effects to place constraints on fundamental physics parameters from the Epoch of Reionization, which may be possible with cross-correlation synergies between 21 cm and galaxy surveys (e.g., Bacon et al. 2015). 
 
\end{itemize}

 To summarize, the availability of rich datasets related to neutral hydrogen evolution in the post-reionization universe will lead to exquisite constraints on galaxy formation and evolution. They also hold the promise of uncovering cosmological and fundamental physics parameters from these exciting epochs.

 \textbf{Acknowledgements:} My research work is supported by the Tomalla Foundation. I thank my collaborators Alexandre Refregier, Adam Amara, Girish Kulkarni, Tirthankar Roy Choudhury with whom most of the work reported here was done, and several others for many useful and interesting discussions. I thank the organizers of the IAUS 333 for a productive and enriching meeting.

\end{document}